\documentclass[aps,12pt,nofootinbib]{revtex4}

\usepackage{graphicx,graphics}
\usepackage{enumerate}
\usepackage{subfigure}
\usepackage{amsmath}
\usepackage{amsfonts}
\usepackage{amssymb}
\usepackage{amsthm}
\usepackage{psfrag}

\catcode`\@=11
\def\numberbysection{\@addtoreset{equation}{section}
         \def\theequation{\arabic{section}.\arabic{equation}}}
\numberbysection

\def\a{\alpha}
\def\d{{\rm d}}
\def\ci{{\rm i}}
\def\b{\beta}

\def\e{\varepsilon}

\begin{document}

\title{Correlations in the $n\rightarrow 0$ limit of the dense $O(n)$ loop model}

\author{V.S. Poghosyan$^1$ and V.B. Priezzhev$^2$}
\affiliation{
$^1$Institute for Informatics and Automation Problems\\ NAS of Armenia, 0014 Yerevan, Armenia\\
$^2$Bogoliubov Laboratory of Theoretical Physics,\\ Joint Institute for Nuclear Research, 141980 Dubna, Russia
}

\begin{abstract}
The two-dimensional dense  $O(n)$ loop model for $n=1$ is equivalent to the bond
percolation and for $n=0$ to the dense polymers or spanning trees. We consider
the boundary correlations on the half space and calculate the probability $P_b$
that a cluster of bonds has a single common point with the boundary.
In the limit $n\rightarrow 0$, we find an analytical expression for $P_b$
using the generalized Kirchhoff theorem.
\end{abstract}

\maketitle


\section{Introduction}
\begin{figure}[!ht]
\includegraphics[width=60mm]{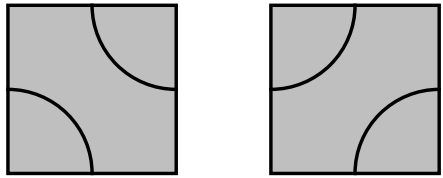}
\caption{\label{fig-cells} Elementary cells.}
\end{figure}

The dense $O(n)$ loop model \cite{dense} is defined by drawing two lines in each
elementary cell of the square lattice. Two possible states of the cell are shown
in Fig.\ref{fig-cells}. The lines on the whole lattice with appropriate boundary conditions
form a system of closed loops with the Boltzmann weight $n$ ascribed to every loop.

\begin{figure}[!ht]
\includegraphics[width=80mm]{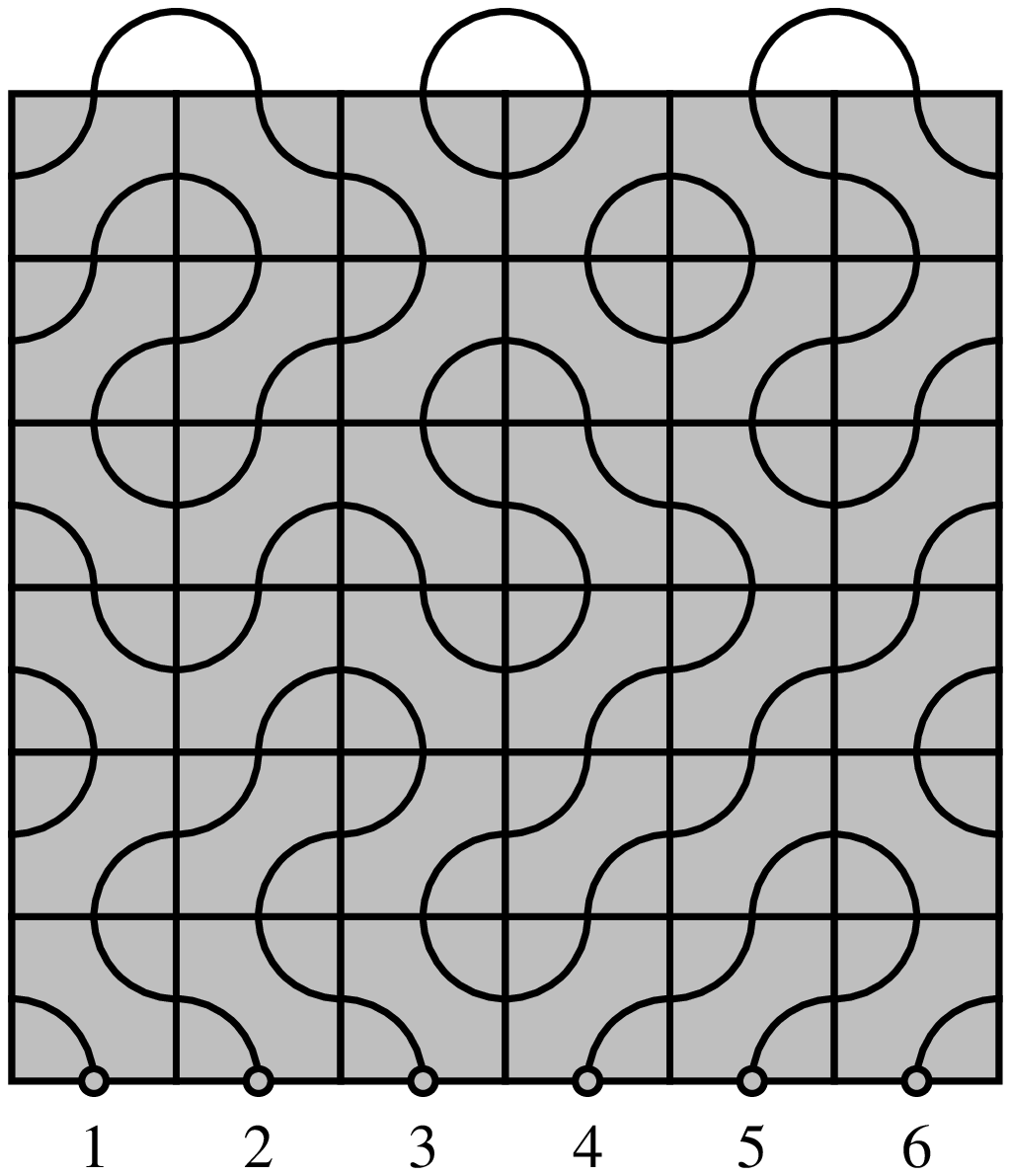}
\caption{\label{fig-loops} Loops on the horizontally periodic lattice.}
\end{figure}

A rise of interest to the $O(n)$ model, especially to the case $n=1$, was triggered by remarkable observations concerning the ground state of the antiferromagnetic XXZ quantum chain and, equivalently, to properties of the transfer matrix of dense loop model taken in the Hamiltonian limit of extreme spatial anisotropy \cite{Stroganov,RazStrog,RS,Batch,deGier,Mitra}. It was noted that the elements of the ground state of the loop Hamiltonian are equal to the cardinality of different subsets of configurations of the Fully Packed Loop (FPL) model which in their turn are connected with classes of Alternating Sign Matrices (ASM).

\begin{figure}[!ht]
\includegraphics[width=80mm]{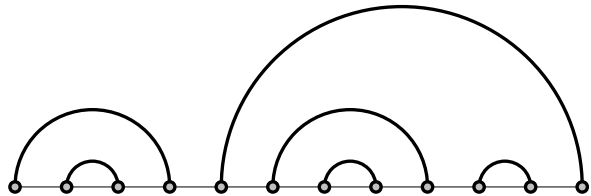}
\caption{\label{fig-links} A link configuration.}
\end{figure}

The states of the $O(n)$ model can be defined in terms of the connectivity condition for points of intersection between loops and a horizontal line cutting the loops at these points. Two points are connected by a link if there exists a line between them via the half space above the cut. For instance, imposing periodic boundary conditions in horizontal direction for the lattice in Fig.\ref{fig-loops} and specifying the boundary conditions at the upper edge, we obtain three minimal links between points 1 and 6, 2 and 3, 4 and 5 which are points of intersection between loops and the bottom line of  lattice belonging to upper half space. A typical configuration of links for a larger lattice is given in Fig.\ref{fig-links}.

Several interesting conjectures have been obtained about the probability distribution of configurations of links of the dense $O(1)$ loop model. To be closer to the aim of this article, we mention just one conjecture about the number of configurations containing the minimal link at a given position \cite{Mitra}
\begin{equation}
\label{even}
\sum  \ldots\frown\ldots =\frac{3}{8}\frac{N^2}{N^2-1} A_{HT}(N)
\end{equation}
for even periodic system and
\begin{equation}
\label{odd}
\sum  \ldots\frown\ldots =\frac{3}{8}\frac{N^2-1}{N^2} A_{HT}(N)
\end{equation}
for odd periodic system. Here $A_{HT}(N)$ is the number of $N\times N$ half
turn invariant ASM \cite{Kuperberg}.

Since the total number of periodic link configurations is $A_{HT}(N)$, the probability of minimal link $Prob(\ldots\frown\ldots)$ tends to $3/8$ in the limit $N \rightarrow \infty$. Due to the simple bijection between the loop configurations on the square lattice and the bond configurations on its sublattices, the dense $O(1)$ model is equivalent to to the bond percolation. Then, the constant $3/8$ has a simple
interpretation as the probability of a certain class of percolation clusters (see Section II). Despite its simplicity, an analytical derivation of the probability
of minimal links as well as of other link configurations remains an open problem.

In this work, we consider the problem of evaluation of $Prob(\ldots\frown\ldots)$
in the dense $O(n)$ model for the case $n=0$. In the limit $n \rightarrow 0$,
the bulk loops disappear and the system of lines is converted into the model of
dense polymers \cite{Pearce}. The bijection giving the bond percolation for $n=1$
on a sublattice, leads in the limit $n \rightarrow 0$ to the spanning trees on the same sublattice. A lot is known about correlation functions of the spanning tree
model due to the Kirchhoff theorem \cite{Kirchhoff}. However, we will see in subsequent sections that evaluation of  $Prob(\ldots\frown\ldots)$ involves non-local diagrams which need a generalization of the  Kirchhoff determinant formula.
Similar correlations appear in the Abelian sandpile theory \cite{Dhar,Pri94,jpr,correlations,exactintegral}, in the problem of loop-erazed random walk \cite{Levine,LERW},
in the problem of ``watermelon'' embedded into the spanning tree \cite{3-leg,watermelon}.
Below, we use the generalized determinant formula to find $Prob(\ldots\frown\ldots)$ and compare the result obtained for $n=0$ with that conjectured for $n=1$.

\section{Link structure and boundary clusters}

\begin{figure}[!ht]
\includegraphics[width=80mm]{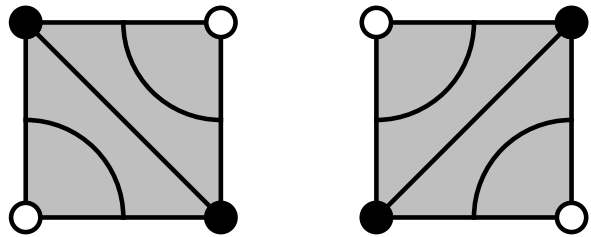}
\caption{\label{fig-bijection} The bijection between bonds and elementary cells.}
\end{figure}

The square lattice of sites with integer coordinates $(n,m)$ can be divided
into two sublattices, black and white. For sites of the black sublattice $n+m$ is
even, for sites of the white one $n+m$ is odd. The bijection between loop and bond configurations mentioned in Introduction is shown in Fig.\ref{fig-bijection}.
The neighboring sites of each sublattice are connected by a bond if it does not intersect the lines of elementary cell. Each connected cluster of bonds in the bulk of lattice is situated
inside a loop. Each bulk cluster on the black sublattice is surrounded by a connected cluster of bonds on the white sublattice and vice versa.

\begin{figure}[!ht]
\includegraphics[width=80mm]{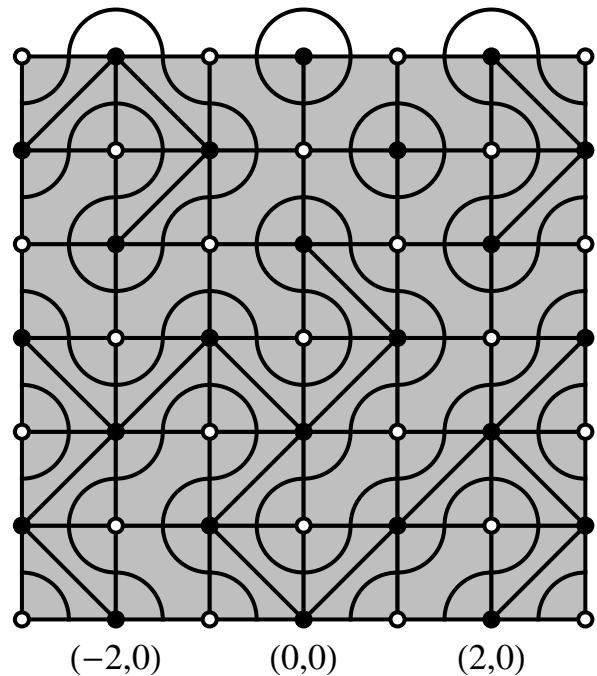}
\caption{\label{fig-loop-bonds} The bond configuration corresponding to the loop configuration from Fig.\ref{fig-loops}.}
\end{figure}

The horizontal line  cutting the loops cuts bond clusters inside them and produces
boundary clusters surrounded by pieces of loops. Every piece has two points of intersections with the horizontal line and corresponds to the link between these
points. The clusters of bonds corresponding to the loop configuration in Fig.\ref{fig-loops} are
shown in Fig.\ref{fig-loop-bonds}.

\begin{figure}[!ht]
\includegraphics[width=100mm]{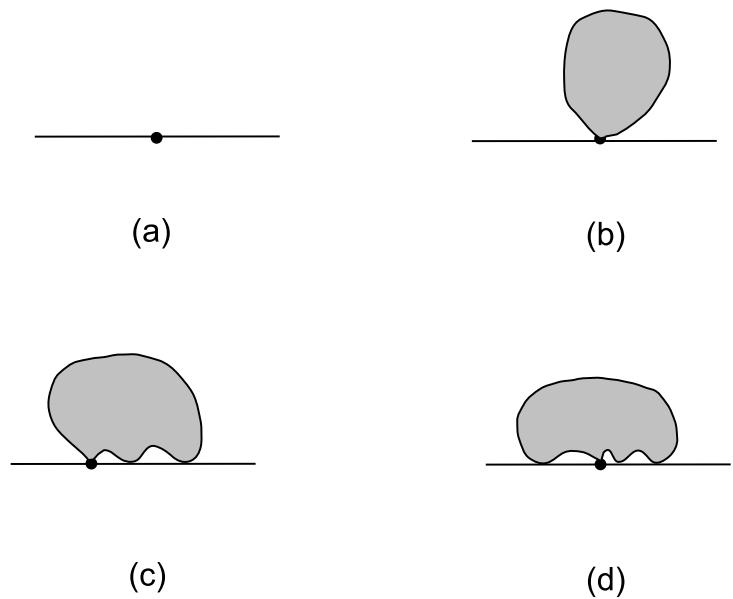}
\caption{\label{fig-boundary-clusters} Four types of the boundary clusters.}
\end{figure}

Since we are interested in the limiting case of large lattice $N \rightarrow \infty$, we take the infinite horizontal line and neglect the left and right boundary conditions.
We put the origin of the lattice at a selected black site on the horizontal line and
consider the boundary clusters of bonds containing the origin.
The boundary clusters  can be classified into four types shown schematically in Fig.\ref{fig-boundary-clusters}.
The cluster of type (a) consists of a single vertex coinciding with the origin.
The type (b) represents clusters containing no points of the horizontal line besides the origin.
The clusters of type (c) contain the origin and one or more boundary vertices left or right from the origin.
The clusters of type (d) contain the origin and an arbitrary number of boundary vertices left and right from the origin.

The minimal link probability $Prob(\ldots\frown\ldots)$ is the sum of probabilities
$P_a$ of the clusters of type (a) and $P_b$ of the clusters of type (b) because the minimal link corresponds to the piece of loop surrounding the origin $(0,0)$ together with the cluster of bonds attached to the origin.

In the $O(1)$ model, the probability $P_a$ is elementary because the isolated site $(0,0)$ of the black sublattice corresponds to two bonds on the dual white sublattice: one bond connecting sites $(-1,0)$ and $(0,1)$ and another one connecting sites $(0,1)$ and $(1,0)$. For the bond percolation, where the bond probability
is $1/2$, $P_a = 1/4$. The conjectured value of  $Prob(\ldots\frown\ldots)$ is $3/8$,
then $P_b=1/8$. Considering the system of links as a height profile \cite{deGier},
one can note that the minimal links correspond to peaks of the profile and the local link configurations produced by clusters of type (d) correspond to valleys of the
profile. The numbers of peaks and valleys in the periodic system coincide, so we have
$P_d=3/8$. Using the identity
\begin{equation}
\label{unity}
P_a + P_b +2P_c + P_d = 1
\end{equation}
we conclude that $P_c = 1/8$. The symmetry $P_b = P_c$ is wonderful because the bond
clusters of type (b),(c) do not obey any visible symmetry in the percolation theory.

\begin{figure}[!ht]
  \includegraphics[width=120mm]{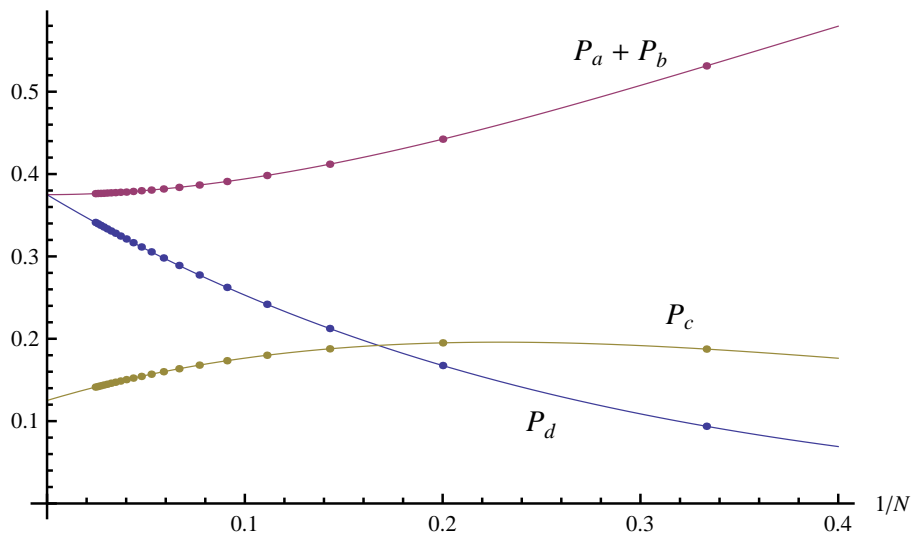}
  \caption{ Monte-Carlo simulation result. Number of samples is $10^8$, the $N \times N$ upper-left half of square with $N=3,5,7,\ldots,41$ is considered.}
  \label{fig-PXY-numerical}
\end{figure}

In Table 1, we show the results of enumeration of clusters of type (a),(b),(c) and (d). We considered the
percolation problem on the upper-left half of square of size $N\times N$ crossed by the diagonal line, so that
the number of lattice points on the vertical, horizontal and diagonal edges of the obtained triangle is $N$.
The origin is put into the central site on the diagonal. To provide a symmetry with respect to the origin, we
take $N$ odd. Despite the difference between our triangle geometry and the strip geometry used in \cite{Mitra},
we see from
Fig.\ref{fig-PXY-numerical} that the convergence of $P_a+P_b$ to $3/8$ is of an order $1/N^2$ as it is expected from the conjectures (\ref{even}) and (\ref{odd}).


\begin{table}
\begin{tabular}{|c|c|c|c|}
  \hline
 $\quad N \quad$ & $\quad\quad P_d \quad\quad$ & $\quad\quad P_a + P_b \quad\quad$ & $\quad P_c \quad$ \\
  \hline\hline

 3 &  0.09374714 & 0.53131273 & 0.18747400 \\ \hline
 5 &  0.16750925 & 0.44235040 & 0.19503558 \\ \hline
 7 &  0.21246768 & 0.41194524 & 0.18778562 \\ \hline
 9 &  0.24184903 & 0.39823148 & 0.17995654 \\ \hline
 11 & 0.26237808 & 0.39096527 & 0.17334548 \\ \hline
 13 & 0.27745483 & 0.38667526 & 0.16799024 \\ \hline
 15 & 0.28900901 & 0.38382077 & 0.16360972 \\ \hline
 17 & 0.29812103 & 0.38188281 & 0.15998131 \\ \hline
 19 & 0.30552463 & 0.38055439 & 0.15694691 \\ \hline
 21 & 0.31146670 & 0.37975536 & 0.15428928 \\ \hline
 23 & 0.31665257 & 0.37888834 & 0.15225164 \\ \hline
 25 & 0.32118728 & 0.37810530 & 0.15038940 \\ \hline
 27 & 0.32475414 & 0.37786765 & 0.14867657 \\ \hline
 29 & 0.32815029 & 0.37744192 & 0.14722638 \\ \hline
 31 & 0.33086748 & 0.37718018 & 0.14601543 \\ \hline
 33 & 0.33341752 & 0.37693901 & 0.14478432 \\ \hline
 35 & 0.33571270 & 0.37664402 & 0.14381127 \\ \hline
 37 & 0.33771167 & 0.37651494 & 0.14288661 \\ \hline
 39 & 0.33955803 & 0.37642572 & 0.14202344 \\ \hline
 41 & 0.34123656 & 0.37621796 & 0.14131214 \\ \hline

 $\infty$ & $0.37497 \approx 3/8$ & $0.37499 \approx 3/8$ & $0.12506 \approx 1/8$ \\ \hline
\end{tabular}
\caption{The dense $O(1)$ loop model: values for $N=\infty$ are obtained from the extrapolation (see Fig.\ref{fig-PXY-numerical}).}
\end{table}


Our aim in this work is calculation of probabilities $P_a,\; P_b,\; P_c,\; P_d$ for the free-fermion $O(0)$ model. Elimination of loops in the limit $n \rightarrow 0$
converts the wavy lines of the $O(n)$ model into dense polymers \cite{Pearce} and
the bond clusters on the black and white sublattices into spanning trees on these
sublattices. A spanning tree on the black (white) sublattice is a connected cluster containing all black (white) sites.
Due to the cut, the connected spanning tree splits into separated components attached to the horizontal line.
The classification of the components (Fig.\ref{fig-boundary-clusters}) is the same as in the percolation model.
For definiteness, we consider black components and calculate their statistics via probabilities of white components surrounding black clusters.
Normally, one selects an arbitrary site of the tree as a root. Then all bonds of the tree become oriented towards the root.
It is convenient to choose the root of tree on the white sublattice at a site below the horizontal line that cuts the lattice.
Then a white cluster having $k$ sites on the cut can be oriented by $k$ different ways because every boundary site can be connected
with the root by a path along the tree.
Evaluation of $P_a$ is almost as elementary as that for the $O(1)$ case. Indeed,
we have to find the probabilities of three configurations of two oriented bonds
on the dual white sublattice:

1a. The bonds from $(-1,0)$ to $(0,1)$ and from $(0,1)$ to $(1,0)$;

2a. The bonds from $(1,0)$ to $(0,1)$ and from $(0,1)$ to $(-1,0)$;

3a. The bonds from $(-1,0)$ to $(0,1)$ and from $(1,0)$ to $(0,1)$.

The probability $P_a$ is the sum of these probabilities, $P_a=P_{1a}+P_{2a}+P_{3a}$.

Evaluation of $P_b$ is a more delicate problem.
The cluster of type (b) on the black sublattice is surrounded by bonds of a cluster
on the white sublattice. The bonds are directed and their sequence gives continuous
paths on the white sublattice of three different forms:

1b. The self-avoiding path of an arbitrary length which starts at the site $(-1,0)$ and stops at $(1,0)$.

2b. The self-avoiding path starts at the site $(1,0)$ and stops at $(-1,0)$.

3b. Two self-avoiding paths start at the sites $(-1,0)$ and $(1,0)$, meet at a white
site $s_0$ and then the single self-avoiding path continues from
$s_0$ up to the end at a white site on the cut. In particular cases, the site $s_0$ can coincide with sites $(-1,0)$ and $(1,0)$.

We denote the probabilities of these situations by $P_{1b}, P_{2b}, P_{3b}$,
and their sum by $P_b=P_{1b}+P_{2b}+P_{3b}-P_a$.

In a similar way, the probabilities $P_c$ and $P_d$ can be defined.
Taking into account that $P_c=P_c(left)=P_c(right)$ due to the symmetry,
we can define  $P_c(left)$ via  probability of paths on the white
sublattice which separate the origin $(0,0)$ from the boundary
vertices with positive even coordinates  $(2n,0), n \geq 1$.
However, we do not need a separate evaluation of $P_c$ because it can be
determined  from the relation $ P_a + P_b = P_d$ and condition
({\ref{unity}}).

While the probabilities constituting $P_a$ are local correlation functions and
can be calculated within a usual approach of the perturbed Laplacian, the
probabilities in $P_b$ are essentially non-local and their calculation needs a
special technique. Before formulating these technical tools, we chose more convenient
coordinates for our problem.

\begin{figure}[!ht]
\includegraphics[width=80mm]{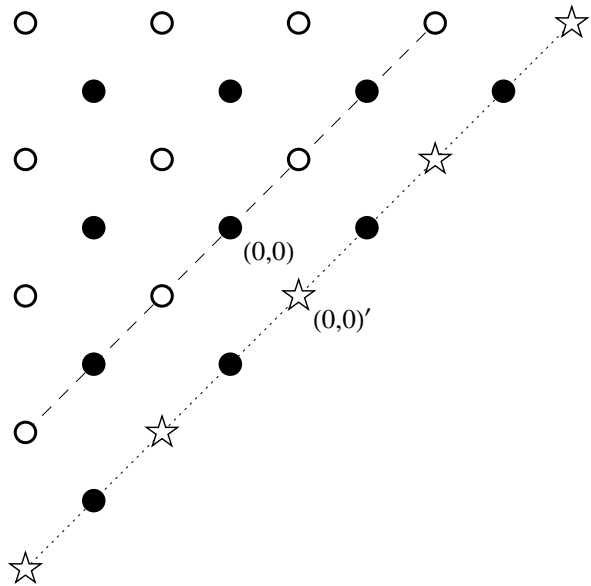}
\caption{\label{fig-upper-left-half-plane} The upper-left half-plane. Open stars are roots of the spanning trees on the white sublattice.}
\end{figure}

Both black and white sublattices are square lattices with the elementary cell
$\sqrt{2} \times \sqrt{2}$ that are turned by $\pi/4$ with respect to the basic lattice. We chose the standard orientation for the sublattices and put the step of
these square lattices to be equal 1. In new coordinates, the former cut is the
diagonal line crossing the origin. The sites of black sublattice belonging to
cut have coordinates $(0,0),\;(1,1),\;(-1,-1),\;(2,2),\;(-2,-2),\dots\;$.
The coordinates of white sublattice on the cut are half-integers (see Fig.\ref{fig-upper-left-half-plane}).
To avoid complication of notations, we shift the origin of the white sublattice to the point $(0,0)'$ shown in Fig.\ref{fig-upper-left-half-plane}.

After the transformation, the spanning trees on the sublattices belong to the upper-left half-space with respect to the diagonal cut.  To describe the structure of roots of the obtained spanning trees, we use the triangular geometry
of the upper-left half of a square exploited above in the numerical simulations of the percolation $O(0)$-model.
Starting with the square of finite size $N\times N$ with closed boundary conditions, we put a single root at the right
lower corner of the square. Then, all branches of the one-component spanning tree on the square are oriented
towards the root. The diagonal cuts the spanning tree and creates a forest of trees on the upper-left half of a square, having an individual root each. We can put all the roots at the line below the cut (the line crossing point $(0,0)^{'}$) since these roots are connected with the single root of the tree on the whole $N\times N$ lattice.
Tending $N$ to infinity, we obtain the structure of roots in our problem.

\section{Laplacian and generalized Kirchhoff theorem}

Consider the graph $\mathcal{G} =(V,E)$ with vertex set
$V$ and set of bonds $E$. The vertices are sites of the square
lattice and an additional point which is the root ``$\star$'': $V\equiv \{s_{x,y}, (x,y) \in \mathbb{Z}^2, |x|\leq M, |y|\leq N \}\cup\{\star\}$. The bonds of $E$ connect neighboring sites of the lattice. We set the condition that all right and bottom boundary vertices $\{s_{x,+ N}\}$ and $\{s_{- M, y}\}$ are connected with the root $\star$. The graph $\mathcal{G}$ represents the finite square lattice of size $(2N+1) \times (2M+1)$.
We consider also the left-upper half plane $V_{+}=\{s_{x,y}, x \leq y\}$ with open boundary conditions at the diagonal sites $V_{0}=\{s_{x,y}, x=y\}$.

We construct the  spanning tree  by using the arrow
representation. To this end, we attach to each vertex $i\in V\backslash \{\star\}$ an arrow directed along one of bonds $(i,i')\in E$
incident to it. Each arrow defines a directed bond $(i\rightarrow i')$ and
each configuration of arrows defines a spanning
directed graph  with set of bonds $\{(i\rightarrow i'): i\in V\backslash\{\star\},\ i'\in V,\ (i,i')\in E\}$.
A sequence of directed bonds $(i_1,i_2),\;(i_2,i_3),\;(i_3,i_4),\dots,\;(i_{m-1},i_m)$
is called the path of length $m$ from the site $i_1$ to the site $i_m$.
This path forms a loop if $i_m=i_1$. No arrows are attached to the root $\star$, so that it is a sink of directed paths. Spanning tree is a spanning digraph without any loops. Our aim is to construct spanning trees, with a prescribed configuration of paths between fixed vertices. These configurations will be investigated by means of the determinant expansion of the Laplace matrix \cite{jpr, Dhar, Pri94}.

Let the vertices of the set $V$, be labeled in arbitrary order from $1$ to $n=|V\backslash\{\star\}|=(2M+1)(2N+1)$. Then Laplacian $\Delta$ of size $n \times n$ has the elements:
\begin{equation}
\label{Delta}
\Delta_{ij} =
\begin{cases}
  z_i & \text{if $i=j$},\\
 -1 & \text{if $i,j$ are nearest neighbors}, \\
  0 & \text{otherwise},
\end{cases}
\end{equation}
where $z_i$ is the degree of vertex $i \in V\backslash \{\star\}$. The determinant of $\Delta$ is a sum over all permutations $\sigma$ of the set $\{1,2,\ldots,n\}$:
\begin{eqnarray}
\label{Leibniz}
\det\Delta=\sum_{\sigma \in S_n} \mathrm{sgn}(\sigma)\Delta_{1,\sigma(1)}\Delta_{2,\sigma(2)}\ldots\Delta_{n,\sigma(n)} \, ,
\end{eqnarray}
where $S_n$ is the symmetric group, $\mathrm{sgn}(\sigma)=\pm 1$ is the signature of permutation $\sigma$. In general, each permutation $\sigma \in S_n$ can be factorized into a composition of disjoint cyclic permutations,  $c_1, c_2, \ldots c_k$. This representation partitions the set of vertices $V\backslash\{\star\}$ into non-empty disjoint subsets which are orbits $\mathcal{O}_i=\{v_{i,1},v_{i,2}\ldots,v_{i,l_i}\}\subset V$ of the corresponding cycles $c_i$, $i=1,\ldots,k$, at that $\cup_{i=1}^k\mathcal{O}_i=V\backslash\{\star\}$ and $\sum_{i=1}^k l_i=n$, where $l_i$ is the length of cycle $c_i$. Orbits consisting of just one element are the fixed points $S_{fp}(\sigma)$ of the permutation. A cycle $c_i$ of length $|c_i|=l_i\geq 2$ is called a proper cycle. The proper cycles are of even length only, hence, the number of proper cycles $p$ defines the signature of the permutation $\sigma$, that is $\mathrm{sgn}(\sigma)=(-1)^p$.
Thus (\ref{Leibniz}) can be written as follows:
\begin{eqnarray}
\label{Leibniz2}
\det\Delta=\prod_{i=1}^n z_i + \sum_{p=1}^{[n/2]} (-1)^p \sum_{\sigma =\{c_1, \ldots, c_p\}}\prod_{i=1}^{p}\Delta_{v_i,c_i(v_i)}\Delta_{c_i(v_i),c_i^2(v_i)}\ldots \Delta_{c_i^{l_i-1}(v_i),v_i}\prod_{j \in S_{fp}(\sigma)}z_j \, ,
\end{eqnarray}
where $c^k_i$ is the $k$-fold composition of the cyclic permutation $c_i$ of even length $l_i$, $v_i \in \mathcal{O}_i(\sigma)$, so that  $c_i^{k-1}(v_i)\neq c_i^k(v_i)$ and $c_{i}^{l_i}(v_i)=v_i$.
The term $\prod_{i=1}^n z_i$ equals to the number of all spanning digraphs having the root $\star$. Each of other terms on the right-hand side of (\ref{Leibniz2}) having a non-zero set of fixed points  equals  up to a sign to $\prod_{j \in S_{fp}(\sigma)}z_j$, because all non-diagonal elements equal to $-1$. That product represents the number of distinct spanning digraphs which have in common the specified cycles $c_1,\ldots,c_p$, and differ in the oriented edges outgoing from vertices $j \in S_{fp}(\sigma)$.
The expansion (\ref{Leibniz2}) can be interpreted in form of the inclusion-exclusion principle \cite{Pr85}. Let $c_1,c_2,\ldots,c_m$ be the list of all possible proper cycles. We define $A_i$, $i=1,2,\ldots,m$ as the set of all spanning digraphs containing the particular cycle $c_i$ and $A_0$ is the set of all spanning digraphs. Let $A_{ST}$ be the set of spanning trees i.e. the set of spanning digraphs containing no cycles.  Then we can write down (\ref{Leibniz2}) in the form:
\begin{eqnarray}
\label{Leibniz3}
\det\Delta=|A_{ST}|=|A_0|-\sum_{i=1}^m |A_i|+\sum_{1\leq i < j \leq m} |A_i\cap A_j|+\ldots+(-1)^m|A_1\cap\ldots\cap A_m| \, ,
\end{eqnarray}
where $|A|$ is cardinality of the set A. (\ref{Leibniz3}) is the Kirchhoff theorem for the number of spanning trees of a given graph \cite{Pr85}.

Now we modify the Laplace matrix changing some non-diagonal elements:
\begin{equation}
\label{DeltaPrime}
\Delta_{ij}' =
\begin{cases}
  z_i & \text{if $i=j$},\\
 -1 & \text{if $i,j$ are nearest neighbors}, \\
-\e & \text{if $(i,j)\in \mathcal{B}\equiv\{(i_1,j_1),\ (i_2,j_2), \ldots (i_r,j_r)$}\}, \\
  0 & \text{otherwise},
\end{cases}
\end{equation}
where $\mathcal{B}$ is some arbitrarily chosen set of $r$ directed bonds (bridges) that are not necessary nearest neighbors on the square lattice.
The determinant expansion (\ref{Leibniz2}) with $\Delta^{'}$ generalizes the
Kirchhoff theorem \cite{Pri94}. The product $\Delta_{i_1,j_1}^{'}\dots \Delta_{i_r,j_r}^{'}=(-\e)^r$  survives in the limit $\lim_{\e \rightarrow \infty}  \det\Delta'/ \e^r$. Permutations corresponding to these terms, contain cycles  with directed bonds $\mathcal{B}$.
If the configuration of bonds $\mathcal{B}$ excludes cycles containing more than one bond from the set $\mathcal{B}$,
the expression $\lim_{\e \rightarrow \infty}  \det\Delta'/ (-\e)^r$ gives the number of configurations with following features: (i) each configuration is the $(r+1)$- component spanning graph; (ii) $k$-th component consists of the path from site $j_k$ to site $i_k$, $1 \leq k \leq r$, and branches of the spanning tree attached to these paths; (iii) $(r+1)$-th component is the spanning tree containing the root $\star$. The ratio of numbers of such configurations and all one-component spanning trees is:
\begin{equation}
\label{DetDelta}
\lim_{\e \rightarrow
\infty}\frac{\det\Delta'}{(-\e)^r\det\Delta}=\lim_{\e
\rightarrow \infty}\frac{\det(I+\delta G )}{(-\e)^r} \, ,
\end{equation}
where $\delta=\Delta'-\Delta$ is the defect matrix and
$G=\Delta^{-1}$ is the
Green function on the plane, which has an explicit integral representation in thermodynamical limit $M,N\rightarrow \infty$ \cite{spitz} (see Appendix).

If $\mathcal{B}$ includes $m$ bonds $(i \rightarrow i^{'})$ between nearest neighbors
on the square lattice, the diagonal elements $\Delta^{'}_{i,i}$ are changed from
$z_i$ to $\e$ and non-diagonal ones  $\Delta^{'}_{i,i^{'}}$ from $-1$ to
$-\e$. In this case, the limit $\lim_{\e \rightarrow \infty}  \det\Delta{'}/ \e^m$ enumerates the spanning trees with an obligatory presence of bonds  $(i \rightarrow i^{'})$.

\section{Cluster probabilities}

Upon above preparations, we are ready to calculate probabilities
$P_a=P_{1a}+P_{2a}+P_{3a}$ and $P_b=P_{1b}+P_{2b}+P_{3b}-P_a$.

\subsection{The probability $P_a$}

\begin{figure}[!ht]
  \includegraphics[height=60mm]{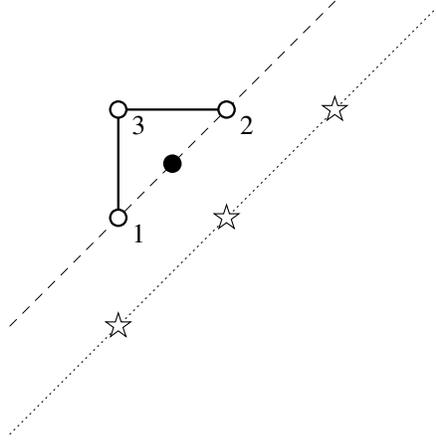}
  \caption{Two-bond configuration. The points 1, 2 and 3 have correspondingly the coordinates $(-1,0)$, $(0,1)$ and $(-1,1)$.}
  \label{fig-minimal-loop}
\end{figure}

The two-bond configuration corresponding to $P_a$ is shown in Fig.\ref{fig-minimal-loop}. We denote by $B$ the non-zero submatrix of the defect matrix $\delta$.
For the case $1a$, matrix $B$ is
\begin{equation}
B(\{1\to 3,3\to 2\})=\left(
\begin{array}{ccc}
\e  & 0 & -\e  \\
0 & 0 & 0 \\
0 & -\e  & \e
\end{array}
\right).
\end{equation}
The probability $P_{1a}$ follows from (\ref{DetDelta})
\begin{equation}
P_{1a}=\lim_{\e
\rightarrow \infty}\frac{\det(I+B G^{\mathrm{op}} )}{(-\e)^2}\,,
\end{equation}
where we changed $\delta$ by its non-zero finite submatrix $B$ and the bulk Green function $G$ by the Green function $G^{\mathrm{op}}$ for
the open diagonal boundary conditions
\begin{equation}
G_{(p_1,q_1),(p_2,q_2)}^{\mathrm{op}} = G_{p_2-p_1,\,q_2-q_1} - G_{p_2-q_1,\,q_2-p_1}\,,
\label{Bond-GF}
\end{equation}
where $(p_1,q_1)$ and $(p_2,q_2)$ are points on the white sublattice with the primed coordinates (see Fig.\ref{fig-upper-left-half-plane}).
This Green function is obtained as a solution of the discrete Poisson equation
\begin{equation}
\Delta G^{\mathrm{op}} = I
\end{equation}
on upper-left half-plane with additional condition that the Green function vanishes on the diagonal line $p=q$ (open boundary).
The boundary Green function (\ref{Bond-GF}) can be easily obtained due the principle of reflection symmetry \cite{Feller}.
Using exact values (\ref{someGF}) for the Green functions at short
distances given in Appendix, we obtain
\begin{equation}
P_{1a} = \frac{ 16\pi -16 -3\pi^2 }{6\pi^2}\,.
\end{equation}
For the case $3a$ we have
\begin{equation}
B(\{1\to 3,2\to 3\})=\left(
\begin{array}{ccc}
\e  & 0 & -\e  \\
0 & \e  & -\e  \\
0 & 0 & 0
\end{array}
\right)
\end{equation}
and
\begin{equation}
P_{3a} = \frac{2 \left( 7\pi -12 -\pi^2 \right)}{\pi^2}\,.
\end{equation}
The resulting probability $P_a$ is
\begin{equation}
P_a = P_{3a} + 2 P_{1a} = \frac{(4 -\pi)(9\pi -22)}{3\pi^2} = 0.181903 \ldots\,
\end{equation}
instead of $P_a=1/4$ for the dense $O(1)$ loop model.

\subsection{The probability $P_b$}

By definition the probability $P_{1b}$ is the probability that the path in a spanning tree starting at site $(-1,0)$
of the original white sublattice stops at site $(1,0)$ and then goes to one of two neighboring roots.

The configuration of this path on the rotated sublattice is shown in

\begin{figure}[!ht]
\includegraphics[height=60mm]{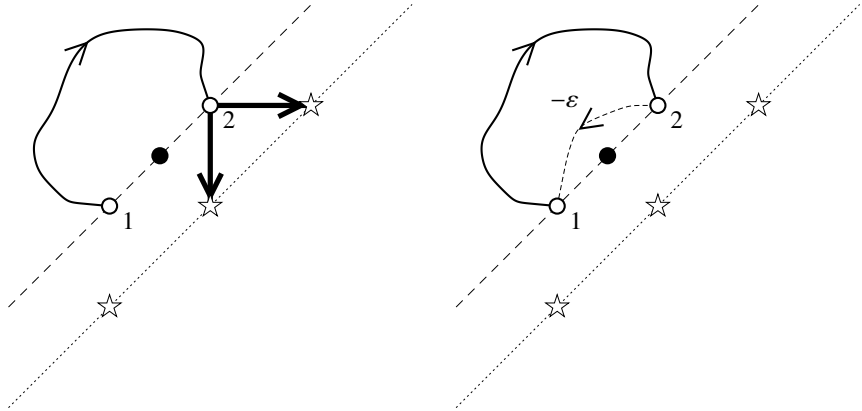}
\caption{ The configuration corresponding to $P_{1b}$ and its bridge representation.
The points 1 and 2 have correspondingly the coordinates $(-1,0)$ and $(0,1)$.}
\label{fig-one-bridge}
\end{figure}
Fig.\ref{fig-one-bridge}, where 1 on the original (non-rotated) lattice is the  site $(-1,0)$ and 2 is the
site $(1,0)$.  On the new lattice their coordinates are correspondingly $(-1,0)$ and $(0,1)$.
Two bold arrows denote two possible connections of  site 2 with the roots.
To generate the path from 1 to 2, we introduce the bridge $(2\rightarrow 1)$ with the weight $-\e$ (see Fig.\ref{fig-one-bridge}) and tend $\e$ to infinity.
Then the defect matrix $B$ consists of the single element $(2 \rightarrow 1)$ and the formula (\ref{DetDelta}) is reduced to
\begin{equation}
\label{short}
P_{1b}=2 \lim_{\e \rightarrow \infty}\frac{\det(I+B G^{op} )}{-\e}=2(G_{1,1}-G_{2,0}) \, .
\end{equation}
Using the table (\ref{someGF}) we get
\begin{equation}
P_{1b}=\frac{2(\pi -3)}{\pi}.
\end{equation}
By the symmetry, we have $P_{1b} = P_{2b}$.

\begin{figure}[!ht]
\includegraphics[width=80mm]{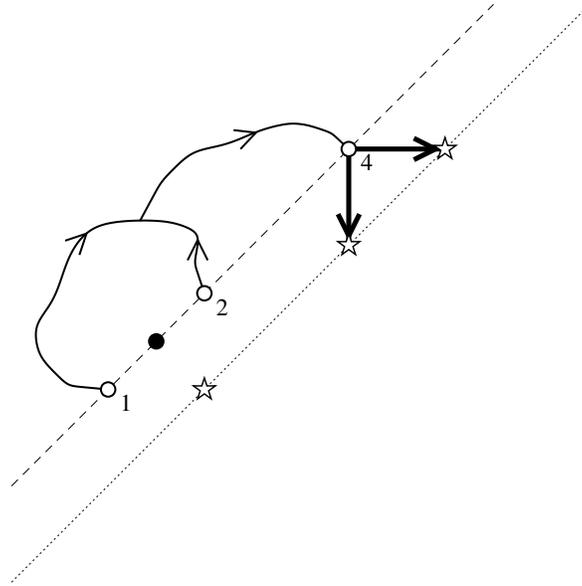}
\caption{\label{fig-diagram-1} The configuration corresponding to case $3b$.}
\end{figure}

The crucial point of our calculations is the probability $P_{3b}$.
The configuration of paths on the white sublattice corresponding to the case $3b$ gets after
the transformation the form depicted in Fig.\ref{fig-diagram-1},
where the point 1 is the transformed position of the site $(-1,0)$, the point 2 of the site $(1,0)$ and the point 4
is the end of a path from the connection point $s_0$. In principle, it is
possible to define a system of auxiliary bridges giving all three paths on
Fig.\ref{fig-diagram-1}, however, in this case we would obtain a product of three
Green functions which should be summed up over all positions of the point
$s_0$. We choose here another way, avoiding these tremendous calculations.

\begin{figure}[!ht]
\includegraphics[width=80mm]{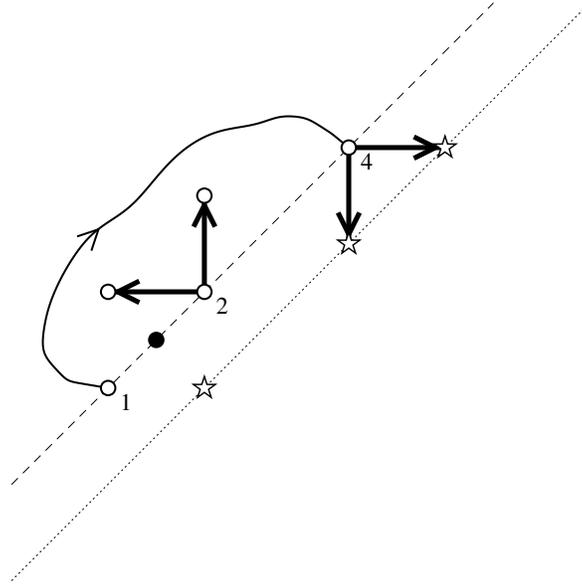}
\caption{\label{fig-diagram-2} The diagram I representing $P_I$.}
\end{figure}

\begin{figure}[!ht]
\includegraphics[width=80mm]{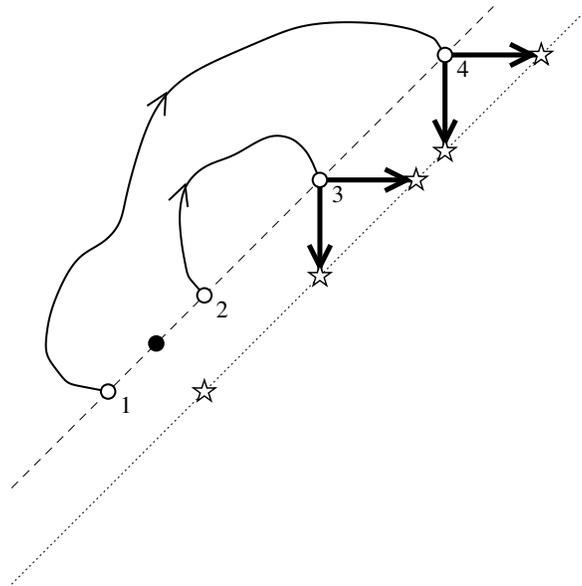}
\caption{\label{fig-diagram-3} The diagram II representing $P_{II}$.}
\end{figure}

First, we define the path from point 1 to point 4 inserting the auxiliary bond $(4\rightarrow 1)$.
Then, we fix an outgoing bond from point 2 in one of two possible directions (diagram I shown in  Fig.\ref{fig-diagram-2}).
The path from point 2 can join the path from point 1 to 4, forming thereby a configuration of the type 3b. Otherwise, the path from 2 can finish at one of the white sites on the boundary between points 2 and 4. The latter case, that is shown on the diagram II,
Fig.\ref{fig-diagram-3}, must be excluded from all configurations in the diagram I.
Thus, instead of summation over two-dimensional positions of the connection point $s_0$, we get two one-dimensional summations over positions of points 3 and 4,
and reduce by one the number of Green functions involved into calculations.

Consider now the Green function  $G_{(p,q),(k,k+1)}^{\mathrm{op}}$ between an arbitrary site $(p,q)$, $q>p$ on the left-upper half part of the square lattice and a site $(k,k+1)$ on the diagonal boundary. The summation over the whole boundary gives
\begin{eqnarray}\label{Gsum}
\sum_{k=-\infty}^{+\infty}G_{(p,q),(k,k+1)}^{\mathrm{op}}=\left.\left.\sum_{k=-\infty}^{+\infty}  \right( G_{k-p,\,k+1-q} - G_{k-q,\,k+1-p} \right)&=& \\
\frac{1}{8\pi^2} \int\!\!\!\!\int_{-\pi}^{\pi}
\sum_{k=-\infty}^{+\infty} \frac{
e^{\ci \, k (\alpha + \beta) } e^{\ci \, \beta} ( e^{ -\ci \, ( p \alpha + q \beta)} - e^{ -\ci \, ( q \alpha + p \beta)} )
}{2-\cos\alpha-\cos\beta}\, \d \alpha \, \d \beta &= &\nonumber \\
= \frac{1}{4\pi} \int_{-\pi}^{\pi} \frac{\ci \, e^{-\ci \, \alpha} \sin[(q-p)\alpha]}{1-\cos\alpha}\, \d \alpha =
  \frac{1}{4\pi} \int_{-\pi}^{\pi} \frac{\sin \alpha \sin[(q-p)\alpha]}{1-\cos\alpha}\, \d \alpha &=& \frac{1}{2}\,.
\end{eqnarray}
With $p=0$ and $q=1$ we have
\begin{equation}
D_{k} \equiv G_{(0,1),(k,k+1)}^{\mathrm{op}} = G_{k,k} - G_{k-1,k+1} \; ,
\end{equation}
and
\begin{equation}
\sum_{k=-\infty}^{+\infty} D_{k} = \frac{1}{2} \; .
\end{equation}
Due to the symmetry relation
\begin{equation}
D_{-k} = D_{k} \; .
\end{equation}
we can find the sum
\begin{equation}
\sum_{k=1}^{+\infty} D_{k} = \frac{1}{4} - \frac{1}{2} (G_{0,0} - G_{1,1}) = \frac{1}{4} - \frac{1}{2\pi} \; .
\end{equation}
Denote by $R_k$ the probability of the diagram I for distance $k$ between
points 1 and 4  (see Fig.\ref{fig-diagram-2}). The defect matrices $B$
containing the auxiliary bond between points 1 and 4, and selected outgoing bonds
from the point 2 is
\begin{equation}
B =\left(
\begin{array}{cccccc}
0 &   0  & 0 & -\e &  0  &  0  \\
0 & -2\e & 0 &  0  & -\e & -\e \\
0 &   0  & 0 &  0  &  0  &  0  \\
0 &   0  & 0 &  0  &  0  &  0  \\
0 &   0  & 0 &  0  &  0  &  0  \\
0 &   0  & 0 &  0  &  0  &  0
\end{array}
\right) \, ,
\end{equation}
where the 5-th and 6-th rows and columns correspond to the two neighbouring vertices of point 2.
The calculation of the determinant
gives
\begin{equation}
R_{k} = \frac{4 (\pi - 3 )}{\pi} D_{k-1} + \frac{2(\pi - 2)}{\pi} D_k \; .
\end{equation}
The probability $P_I$ is the sum
\begin{equation}
P_I = \sum_{k=2}^{\infty} R_{k} \; .
\end{equation}
From the identity
\begin{eqnarray}
\sum_{k=-\infty}^{+\infty} R_{k} = 2 \sum_{k=2}^{+\infty} R_{k} +
\frac{4 (\pi - 3 )}{\pi} D_{0} + \frac{2(\pi - 2)}{\pi} \left( D_{-1} + D_{0} + D_{1} \right) \;
\end{eqnarray}
we have
\begin{equation}
P_I = \frac{6\,\pi - 8 - \pi^2}{2\,\pi^2}.
\end{equation}

The probability $P_{II}$ of the diagram II (see Fig.\ref{fig-diagram-3}) is a double sum of double bridges
with distance $s$ between points 1 and 4 and distance $k$ between points 1 and 3. The corresponding $B$ matrix reads
\begin{equation}
B =\left(
\begin{array}{cccc}
0 & 0 & 0 &-\e \\
0 & 0 &-\e& 0  \\
0 & 0 & 0 & 0  \\
0 & 0 & 0 & 0  \\
\end{array}
\right) \, ,
\end{equation}
which gives the probability
\begin{eqnarray}
P_{II} &=& 4 \lim_{\e \rightarrow \infty}\frac{\det(I+B G^{op} )}{\e^2} = 4 \sum_{2 \leq k < s} ( D_{k - 1}D_{s} - D_{k}D_{s - 1} ) =\\
&=& 4 \sum_{2 \leq k \leq s} D_{s}( D_{k - 1} - D_{k} ) + 4 \sum_{k=2}^{\infty} \sum_{s=k}^{\infty} D_{k}( D_{s} - D_{s-1} ) =\\
&=& 4 \sum_{s=2}^{\infty} D_{s}( D_{1} - D_{s} ) - 4 \sum_{k=2}^{\infty} D_{k} D_{k-1} =\\
&=& 4 D_{1} \sum_{s=2}^{\infty} D_{s} - 4 \sum_{k=2}^{\infty} D_{k}( D_{k} + D_{k-1} )\;,
\end{eqnarray}
or, in a more convenient form,
\begin{equation}
P_{II} = 4 D_1 \sum_{k=1}^{+\infty} D_{k} - 4 \sum_{k=1}^{+\infty} D_{k}( D_{k} + D_{k+1} )\;.
\end{equation}
One can show that
\begin{equation}
\sum_{k=-\infty}^{+\infty} D_{k}( D_{k} + D_{k+1} ) = 2 \sum_{k=1}^{+\infty} D_{k}( D_{k} + D_{k+1} ) + D_0 (D_0 + 2 D_1 )
\end{equation}
so
\begin{equation}
P_{II} = 4 D_1 \sum_{k=1}^{+\infty} D_{k} + 2 D_0 (D_0 + 2 D_1 ) - 2 \sum_{k=-\infty}^{+\infty} D_{k}( D_{k} + D_{k+1} ) \; .
\end{equation}
Denote the latter sum by $Q$:
\begin{equation}
Q = \sum_{k=-\infty}^{+\infty} D_{k}( D_{k} + D_{k+1} )\;.
\end{equation}
Using the integral representation of the Green function and definition of $D_{k}$, after some manipulations we can show that
\begin{equation}
Q = \frac{1}{16\,\pi^3} \int\!\!\!\!\int\!\!\!\!\int_{-\pi}^{\pi}
\frac{( \sin\a_1 -\sin\b_1 ) (\sin\a_1 - \sin( 2\a_1 + 2\a_2 + \b_1 ))}{(2 - \cos\a_1 - \cos\b_1 ) (2 - \cos\a_2 - \cos(\a_1+\b_1+\a_2))}
\, \d\a_1 \, \d\b_1 \, \d\a_2 \, .
\end{equation}
By the substitution
\begin{equation}
t_1 = \tan\frac{\alpha_1}{2}, \quad\quad\quad
t_2 = \tan\frac{\beta_1 }{2}, \quad\quad\quad
t_3 = \tan\frac{\alpha_2}{2},
\end{equation}
\begin{equation}
\sin\a_1=\frac{2 t_1}{1+t_1^2},   \quad
\sin\b_1=\frac{2 t_2}{1+t_2^2},   \quad
\sin\a_2=\frac{2 t_3}{1+t_3^2},
\end{equation}
\begin{equation}
\cos\a_1=\frac{1-t_1^2}{1+t_1^2}, \quad
\cos\b_1=\frac{1-t_2^2}{1+t_2^2}, \quad
\cos\a_2=\frac{1-t_3^2}{1+t_3^2},
\end{equation}
\begin{equation}
\d\a_1 \, \d\b_1 \, \d\a_2 = \frac{2}{1 + t_1^2} \frac{2}{1 + t_2^2} \frac{2}{1 + t_3^2} \, \d t_1 \, \d t_2 \, \d t_3
\end{equation}
and symmetrization by permutation $t_1 \rightleftarrows t_2$, we come to the triple integral of a rational function
\begin{equation}
Q = \frac{1}{\pi^3} \int\!\!\!\!\int\!\!\!\!\int_{-\infty}^{\infty} T_1(t_1,t_2,t_3) T_2(t_1,t_2,t_3) \, \d t_1 \, \d t_2 \, \d t_3 \, ,
\end{equation}
where
\begin{equation}
T_1(t_1,t_2,t_3) =
\frac{ (t_1 - t_2)^2 (t_1 t_2 - 1) (t_1 t_2 + t_1 t_3 + t_2 t_3 -1) (t_1 t_2 t_3 - t_1 - t_2 - t_3) }
{ (1 + t_1^2 )^2 (1 + t_2^2 )^2 (1 + t_3^2 )^2 (t_1^2 + t_2^2 + 2 t_1^2 t_2^2) } \, ,
\end{equation}
\begin{equation}
T_2(t_1,t_2,t_3) =
\frac{2 t_1 t_2 t_3 - 2 t_3 + (t_1 + t_2) (t_3^2 - 1)}
{(t_1 + t_2 + t_3)^2 - 2 t_1 t_2 t_3 (t_1 + t_2 + t_3) + 2 t_1^2 t_2^2 t_3^2 + (1 + t_1^2 + t_2^2)
t_3^2} \, .
\end{equation}
After integration over $t_3$ and several manipulations we have
\begin{equation}
Q = \frac{2}{\pi^2} \int_{0}^{+\infty} \!\!\! \int_{-t_1}^{+t_1}\,
\frac{(t_1-t_2)^2 \left( 1 + t_1^2 + t_2^2 + t_1^2 t_2^2 - (t_1 + t_2) \sqrt{(1 + t_1^2)(1 + t_2^2)} \right)}
{(1 + t_1^2)^2 (1 + t_2^2)^2 ( t_1^2 + t_2^2 + 2 t_1^2 t_2^2 )}
\,\d t_2 \, \d t_1 \, .
\end{equation}
The integration over $t_2$ gives
\begin{equation}
Q = \frac{1}{\pi^2} \int_{0}^{+\infty}
\left(
\frac{8 t_1^3 \arctan\left( \sqrt{1 + 2 t_1^2} \right)}{(1 + t_1^2)^2 \sqrt{1 + 2 t_1^2}}
+\frac{4 (1 - t_1^2) \arctan (t_1)}{(1 + t_1^2)^2}
-\frac{2 (\pi -2) t_1^2}{(1 + t_1^2)^2}
\right) \, \d t_1 \, ,
\end{equation}
from which we obtain
\begin{equation}
Q = \frac{4 - \pi}{2\pi} \, .
\end{equation}
As a result, we have
\begin{equation}
P_{II} = \frac{ 2\pi^2 - 5 \pi - 4 }{\pi^2} \, .
\end{equation}
and
\begin{equation}
P_{3b} = 2 \times P_I - 2 \times P_{II} = \frac{16-5\pi}{\pi} \, .
\end{equation}

The cumulated result is
\begin{equation}
P_{1b} + P_{2b} + P_{3b} = \frac{ 4 - \pi }{\pi} = 0.27323\ldots \, .
\label{P1b+P2b+P3b}
\end{equation}
and we obtain the probability of clusters $P_b$ in the dense $O(0)$ loop model:
\begin{equation}
P_b = P_{1b} + P_{2b} + P_{3b} - P_a = \frac{2 ( 4 - \pi ) ( 11 - 3 \pi )}{3 \pi^2} = 0.091336465\ldots \, .
\end{equation}

The probabilities $P_c$ and $P_d$ can be easily determined from (\ref{unity}) and the condition
$P_a + P_b = P_d$.

\begin{figure}[!ht]
\includegraphics[width=80mm]{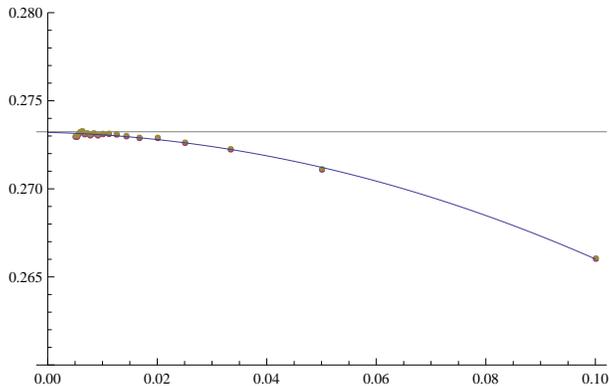}
\caption{ The Monte-Carlo simulation test for (\ref{P1b+P2b+P3b}).
           The dots correspond to $N=10,20,\ldots,200$ and number of simulations $10^8$.}
\label{fig-P1b+P2b+P3b}
\end{figure}

Using the one-to-one correspondence between spanning trees and loop erased random walk (Wilson algorithm, see \cite{br,al,pe,Majumdar,wil} for details),
we can generate equally distributed spanning trees and check the obtained analytical results.
Fig.\ref{fig-P1b+P2b+P3b} with number of samples $10^8$ shows that the numerical value coincides with analytical result (\ref{P1b+P2b+P3b}) with high precision.

The calculation of the minimal link probability for $n=0$ is a first step in
a program of investigations of more general link correlation functions.
Being realized, this program would help to establish a correspondence between lattice calculations
and operators of the logarithmic conformal field theory (LCFT) with
the central charge $c=-2$. Then, the operators of the LCFT corresponding to link
variables in the dense $O(1)$ loop model with the central charge $c=0$ can be
constructed by an analogy with the $O(0)$ model.

\section{Appendix: Green functions}

The explicit form of the translationally invariant Green function on the plane is \cite{spitz}
\begin{equation}
G_{(p_1,q_1),(p_2,q_2)} \equiv G_{\vec{r}_1,\vec{r}_2} \equiv G(\vec{r}_2-\vec{r}_1) \equiv G_{0,0} + g_{p,q}, \quad \vec{r}_2-\vec{r}_1\equiv\vec{r}\equiv(p,q)
\end{equation}

with $G_{0,0}$ an irrelevant infinite constant. The (finite) numbers $g_{p,q}$ are given explicitly by
\begin{equation}
g_{p,q}= \frac{1}{8\pi^2} \int\!\!\!\!\int_{-\pi}^{\pi}
\frac{e^{ \ci\, p \, \alpha + \ci \, q \, \beta}-1}{2-\cos\alpha-\cos\beta} \, \d \alpha \, \d \beta \, .
\label{Green}
\end{equation}
Let us mention symmetry properties of this function:
\begin{equation}
g_{p,q}=g_{q,p}=g_{-p,q}=g_{p,-q} \, .
\label{GreenSymmetry}
\end{equation}
After the integration over $\alpha$, it can be expressed in a more convenient form for actual calculations,
\begin{equation}
g_{p,q} = \frac{1}{4\pi} \int_{-\pi}^{\pi} \frac{t^p \, e^{ \ci\, q\, \beta} - 1 }{ \sqrt{y^2-1} } \,  \d \beta \, ,
\label{Green2}
\end{equation}
where $t = y - \sqrt{y^2-1}$, $y = 2 - \cos{\beta}$ (see \cite{jpr}). For $r^2 = p^2 +q^2 \gg 1$ it has a behavior \cite{3-leg}
\begin{equation}
\label{GreenFunctionAsymptAppendix}
g_{p,q} = -\frac{1}{2\pi}\left(\log r + \gamma +\frac{3}{2}\log 2\right)
+\frac{\cos (4\, \varphi )}{24\, \pi\,  r^2}
+\frac{18 \cos (4\, \varphi )+ 25 \cos (8\, \varphi )}{480\, \pi\, r^4} + \ldots \, ,
\end{equation}
where $(p,q) = (r\cos \varphi,r\sin \varphi)$, and with $\gamma = 0.57721\ldots$ the Euler constant.
Let us also write down some values of the Green functions for particular points $(p,q)$, which are needed for calculations:
\begin{equation}
\begin{array}{lllll}
  g_{0, 1} = - \frac{1}{4}    & \quad & g_{0, 2} = -1 + \frac{2}{\pi}            & \quad & g_{0, 3} = -\frac{17}{4} + \frac{12}{\pi} \\
  g_{1, 1} = - \frac{1}{\pi}  & \quad & g_{1, 2} = \frac{1}{4} - \frac{2}{\pi}   & \quad & g_{1, 3} = 2 - \frac{23}{3\pi} \\
  g_{2, 2} = - \frac{4}{3\pi} & \quad & g_{2, 3} = -\frac{1}{4} - \frac{2}{3\pi} & \quad & g_{3, 3} = - \frac{23}{15\pi} \, .
\end{array}
\label{someGF}
\end{equation}

\section*{Acknowledgments}
This work was supported by the RFBR grants No. 12-01-00242a, 12-02-91333a,
the Heisenberg-Landau program, the DFG grant RI 317/16-1 and the grant of
NRU Higher School of Economics, Academic Fund program No.12-09-0051.
V.B.P. thanks Vladimir Rittenberg for useful discussions.
V.S.P. thanks Philippe Ruelle for hospitality in UCL (Louvain-la-Neuve, Belgium) and critical remarks.
The numerical simulations were performed on Armenian Cluster for High Performance Computation (ArmCluster, www.cluster.am).

\end{document}